\documentclass[11pt,letterpaper]{article}
\usepackage{naaclhlt2015}
\usepackage{times}
\usepackage{latexsym}
\usepackage{amsmath}
\usepackage{graphicx}
\usepackage{amssymb}
\usepackage{multirow}
\usepackage{etoolbox}

\title{Extending a Single-Document Summarizer to Multi-Document: a Hierarchical
	Approach}

\author{Lu\'is Marujo$^{1,2,3}$, Ricardo Ribeiro$^{1,4}$, David Martins de Matos$^{1,2}$, \\
	 \textbf{Jo\~{a}o P. Neto$^{1,2}$, Anatole Gershman$^{3}$, and Jaime Carbonell$^{3}$}\\
$^1$INESC-ID Lisboa, $^2$IST/ULisboa, $^4$ISCTE-IUL, Lisboa,
Portugal\\$^3$School of Computer Science, CMU, Pittsburgh, USA\\
	    {\tt \{lmarujo,anatoleg,jgc\}@cs.cmu.edu } \\
	    {\tt	\{ricardo.ribeiro,david.matos,joao.neto\}@inesc-id.pt }
}

\date{}

\begin{document}
\maketitle
\begin{abstract}
  The increasing amount of online content motivated the development of
  multi-document summarization methods. In this work, we explore
  straightforward approaches to extend single-document summarization
  methods to multi-document summarization.  The proposed methods are
  based on the hierarchical combination of single-document summaries,
  and achieves state of the art results.
\end{abstract}

\section{Introduction}

The use of the Internet to fulfill generic information needs motivated
pioneer multi-document summarization efforts as
NewsInEssence~\cite{radev:et:al:2005} or
Newsblaster~\cite{McKeown:2002}, online since 2001. In general,
multi-document summarization approaches have to address two different
problems: passage selection and information ordering. Current
multi-document systems adopt, for passage selection, approaches similar
to the ones used in single-document summarization, and use the
chronological order of the documents for information
ordering~\cite{christensen:2013}. The problem is that most approaches
fail to generate summaries that cover generic topics which comprehend
different, equally important, subtopics.

We propose to extend a state-of-the-art single-document summarization
method, \textsc{KP-Centrality}~\cite{Ribeiro&Marujo:2013:KP-Centrality},
capable of focusing on diverse important topics while ignoring
unimportant ones, to perform multi-document summarization. We explore
two hierarchical strategies to perform this extension.

% In this work, we also use as a starting point a single document summarization
% method (see~\cite{Ribeiro&Marujo:2013:KP-Centrality}), but concentrate on
% improving the use of temporal information. In that sense, our claim is that on a
% collection of news stories to summarize, the older ones are less important.

This document is organized as follows: Sect.~\ref{sect:RelatedWork} addresses the
related work; Sect.~\ref{multi-doc-summ} presents our multi-document
summarization appproach; experimental results close the paper.

% Note that although considering the chronological order of the input documents,
% we focus on multi-document summarization, and not on the update summarization
% task, where a user has previous information on a topic and receives more
% information about that topic that should be summarized.
%
\section{Related Work}
\label{sect:RelatedWork}

Most of the current work in automatic summarization focuses on
extractive summarization.  The most popular baselines for
multi-document summarization fall into one of the following general
models:
Centrality-based~\cite{radev:et:al:2004,erkan:radev:2004,Wang:2008,ribeiro:matos:2011},
Maximal Marginal Relevance
(MMR)~\cite{carbonell:goldstein:1998,Guo:2010,Sanner:2011,Lim:2012},
and Coverage-base
methods~\cite{lin:hovy:2000,Sipos:2012}. Additionally, methods such as
\textsc{KP-Centrality}~\cite{Ribeiro&Marujo:2013:KP-Centrality}, which
is centrality and coverage-based, follow more than one paradigm. In
general, Centrality-based models are used to produce generic
summaries, while the MMR family generates query-oriented
ones. Coverage-base models produce summaries driven by words, topics or
events.

Centrality-as-relevance methods base the detection of the most salient
passages on the identification of the central passages of the input
source(s). One of the main representatives of this family is
\textit{Passage-to-Centroid Similarity-based
  Centrality}. Centroid-based methods build on the idea of a
pseudo-passage that represents the central topic of the input
source---the \textit{centroid}---selecting as passages to be included
in the summary the ones that are close to the centroid. Another
approach to centrality estimation is to compare each candidate passage
to every other passage and select the ones with higher scores (the
ones that are closer to every other passage): the \textit{Pair-wise
  Passage Similarity-based Centrality}.

MMR~\cite{carbonell:goldstein:1998} is a query driven relevance model
based on the following mathematical model: {\small \begin{equation*}
\arg\max_{S_i}\Big[\lambda(Sim_1(S_i,
Q))-(1-\lambda)(\max_{S_j}Sim_2(S_i, S_j))\Big]
\label{eq:mmr}
\end{equation*}}
\noindent where $Sim1$ and $Sim2$ are similarity metrics that do not have to be
different; $S_i$ are the yet unselected passages and $S_j$ are the
previously selected ones; $Q$ is the required query to apply the
model; and, $\lambda$ is a parameter that allows to configure the
result to be from a standard relevance-ranked list ($\lambda = 1$) to
a maximal diversity ranking ($\lambda = 0$).

Coverage-based summarization defines a set of concepts 
that need to occur in the sentences selected for the summaries.
The concepts are events~\cite{Filatova:2004}, topics~\cite{lin:hovy:2000},
salient words~\cite{Lin:Bilmes:2010,Sipos:2012}, and word n-grams~\cite{Gillick:2008,Almeida:Martins:2013}.

\section{Multi-Document Summarization}
\label{multi-doc-summ}

Our multi-document approach is built upon a centrality and coverage-based
single-document summarization method,
\textsc{KP-Centrality}~\cite{Ribeiro&Marujo:2013:KP-Centrality}. This method,
through the use of key phrases, is easily adaptable and has been shown to be robust in
the presence of noisy input. This is an important aspect considering that using as input several
documents frequently increases the amount of unimportant content).

% In that sense, an important issue to take into consideration is the length of
% the intermediate summaries. A simple approach is to use the size of the final
% summary. Another alternative is to give additional importance to the most recent
% documents, because they commonly include descriptions of the previous events in
% a summarized form. These summaries are iteratively combined until we reach the
% final summary. Thus, the initial size of the summaries is defined by $L_0 = K
% \times log(i + \phi)$, where $K$ is the maximum number of words of the final
% summary, $i$ is the index of the ascending chronologically ordered collection of
% documents, and $\phi$ is a coefficient to avoid zeros (e.g., $\phi = 2$).

% We tested both alternatives in the our preliminary experiments in the
% evaluation datasets (Section \ref{sect:EvalDatasets}). We observed an average
% improvement between 0.3 and 2 ROUGE points.

When adapting a single-document summarization method to perform
multi-document summarization, a possible strategy is to combine the
summaries of each document.  To iteratively combine the summaries, we
explore two different approaches: single-layer hierarchical and
waterfall. Given that the summarization method also uses as input a
set of key phrases, we extract from each input document the required
set of key phrases, join the extracted sets, and rank the key phrases
using their frequency. To generate each summary, we use the top key
phrases, excluding the ones that do not occur in the input document.

\subsection{Single-Document Summarization Method}

To retrieve the most important sentences of an information source, we
used the \textsc{KP-Centrality}
method~\cite{Ribeiro&Marujo:2013:KP-Centrality}.  We chose this model
for its adaptability to different types of information sources (e.g.,
text, audio and video), while supporting privacy
\cite{Marujo:PIR:2014}, and offering state-of-art performance. 
It is based on the notion of combining key phrases with support sets. 
A
support set is a group of the most semantically related passages.
These semantic passages are chosen using heuristics based on the
passage order method~\cite{ribeiro:matos:2011}.  This type of
heuristics uses the structure of the input document (source) to
partition the candidate passages to be included in the support set in
two subsets: the ones closer to the passage associated with the
support set under construction and the ones further apart.  These
heuristics use a permutation, $d^i_1, d^i_2, \cdots, d^i_{N-1}$, of
the distances of the passages $s_k$ to the passage $p_i$, related to
the support set under construction, with
$d^i_k = dist(s_k, p_i)\text{, }1 \le k \le N-1$, where $N$ is the
number of passages, corresponding to the order of occurrence of
passages $s_k$ in the input source. The metric that is normally used
is the cosine distance.

The KP-Centrality method consists of two steps.  First, it extracts
key phrases using a supervised approach~\cite{Marujo_LREC_2012} and
combines them with a bag-of-words model in a compact matrix
representation, given by:
\AtBeginEnvironment{bmatrix}{\setlength{\arraycolsep}{1pt}}
\begin{equation}
	\begin{bmatrix}
		w(t_1,p_1) & \dots & w(t_1,p_N) & w(t_1,k_1)&\dots& w(t_1,k_M)\\
		\vdots & & & & & \vdots\\
		w(t_T,p_1) & \dots & w(t_T,p_N) & w(t_T,k_1)&\dots& w(t_T,k_M)\\
	\end{bmatrix},
	\label{eq:kpcentrality}
\end{equation}
where $w$ is a function of the number of occurrences of term $t_i$ in passage $p_j$ or key phrase $k_l$, $T$ is the number of terms and $M$ is the number of key phrases. 
Then, using a segmented information source $I \triangleq p_1,p_2,\dots,p_N$, a 
support set $S_{i}$ is computed for each passage $p_i$ using:
\begin{equation}
	S_i \triangleq \{s \in I \cup K : sim(s, q_i) > \varepsilon_i \wedge s \neq q_i\},
	\label{eq:kpsupportset}
\end{equation}
for $i=1,\dots, N+M$. Passages are ranked excluding the key phrases $K$ ({\em artificial passages}) according to:
\begin{equation}
	\operatorname*{arg\,max}_{s \in (\cup^{n}_{i=1}S_i)-K} \big|\{S_i: s \in S_i\}\big|.
	\label{eq:kpmodel:centrality}
\end{equation}

\subsection{Single-Layer Hierarchical}

\begin{figure}[h]
	\centering
	\includegraphics[width=0.75\columnwidth]{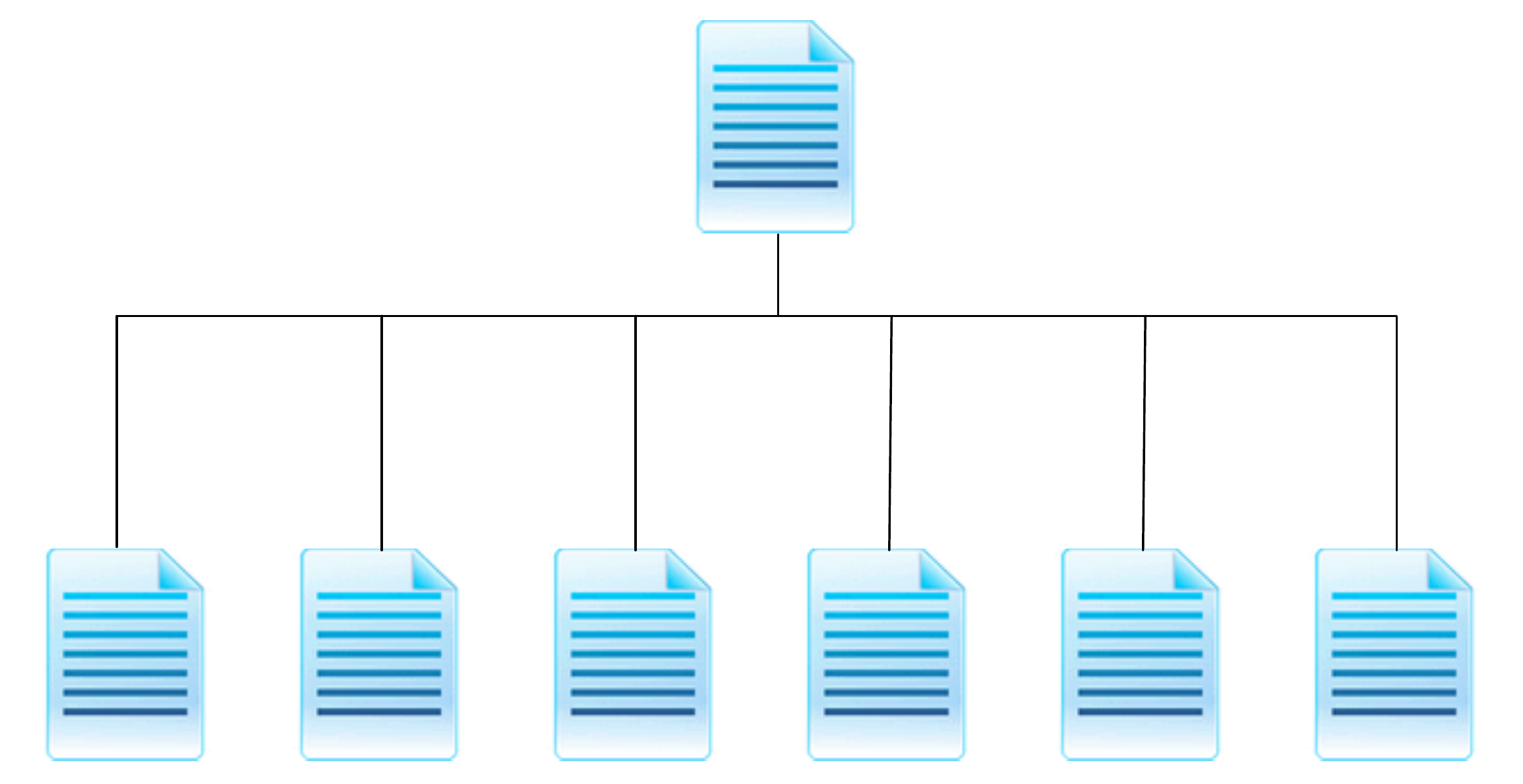}
	\caption{Single-layer architecture.}
	\label{fig:single}
\end{figure}

In this model, we use \textsc{KP-Centrality} to generate, for each news
document, an intermediate summary with the same size of the output summary
for the input documents. An aggregated summary is obtained by
concatenating the chronologically ordered intermediate summaries. The
output summary is again generated by applying \textsc{KP-Centrality} to
the aggregated summary as Figure~\ref{fig:single} shows.

\subsection{Waterfall}

\begin{figure} [h]
	\centering
	\includegraphics[width=0.75\columnwidth]{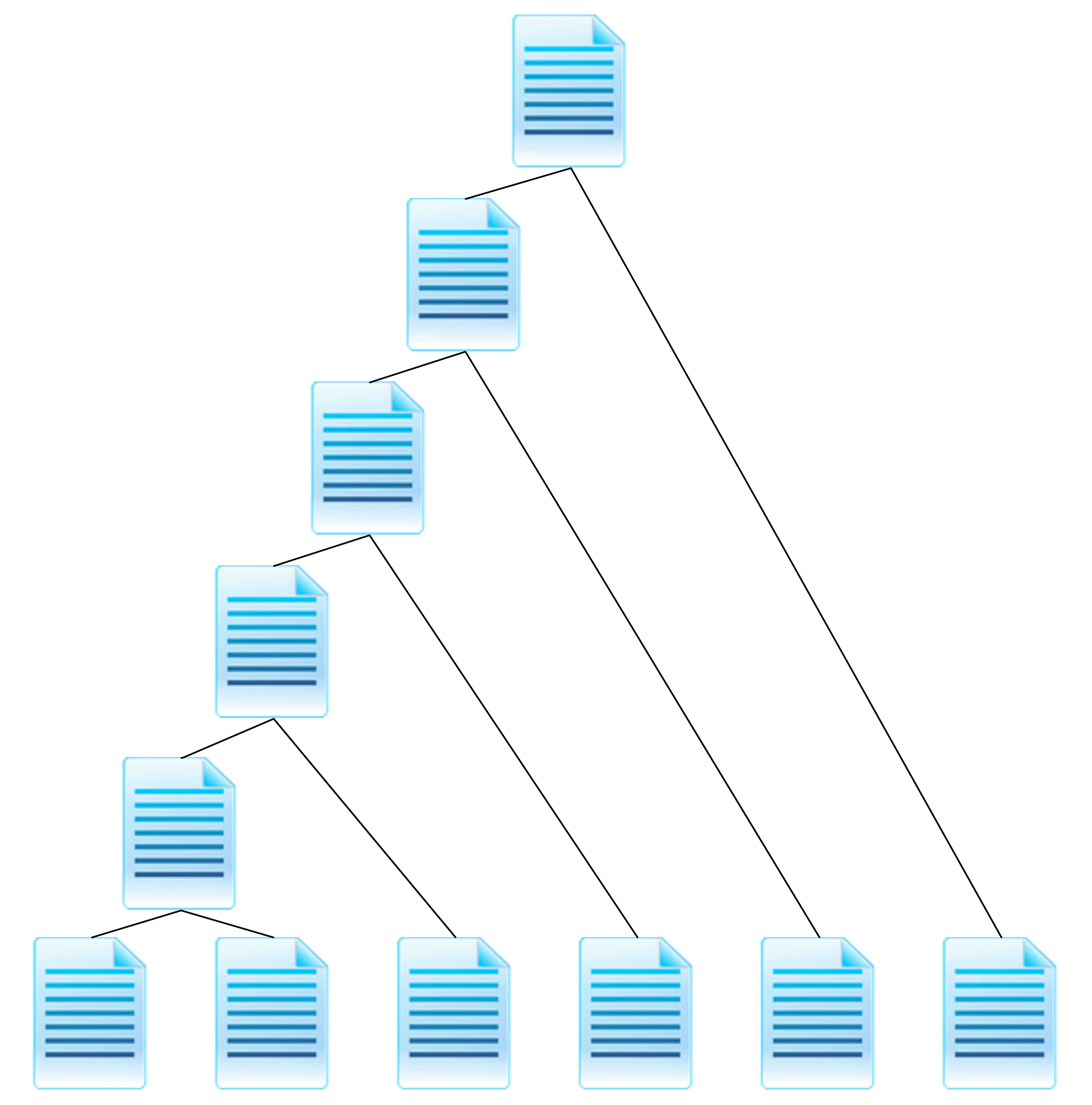}
	\caption{Waterfall architecture.}
	\label{fig:cascade}
\end{figure}

This model differs from the previous one in the
merging process. The underlying merging of the documents follows a
cascaded process: it starts by merging the intermediate summaries, with
the same size of the output summary, of the first two documents, according
to their chronological order. This document is then summarized and merged with
the summary of following document. We iterate this process through all
the documents until the most recent one as Figure~\ref{fig:cascade} illustrates.

% In this model, we use \textsc{KP-Centrality} to generate a summary for
% each news document with size $L_0$. An aggregated summary is obtained
% by concatenating the chronologically ordered summaries. The final
% summary is again generated by applying \textsc{KP-Centrality} to the
% aggregated summary.

% \item[Multilayer Hierarchical]: In this model, we start once again by
% summarizing the individual documents to size $L_0$ (Equation
% \ref{eq:initialSummarySize}). However, the merging of the summaries is performed
% using a hierarchical method, where consecutive documents (in chronological
% order) are paired, merged, and then summarized to size $L_0$. Essentially, we
% are considering each pair of documents is a single-layer. Each one of the
% generated summaries is once again merged until only one final summary is left.

% While the initial size of the summaries of each document is defined by the
% previous equation, for the following summaries we increased their size using a
% scale factor $\delta$, e.g., $\delta = 15$, according to $L_j = \delta \times K
% \times log(i + \phi)$.  Increasing the size of the intermediate summaries is
% fundamental to avoid deleting some important information and to increase some
% redundancy that is helpful to generate the final summary.

%	We manually tried several values for $\delta$ in the evaluation datasets. 
%	As a res 	 

\section{Experimental Results}

We compare the performance of our methods against other representative
models, namely MEAD, MMR, Expected n-call@k~\cite{Lim:2012}, and the
Portfolio Theory~\cite{Wang:2009}. MEAD is a centroid-based
method and one of the most popular centrality-based methods. MMR
is one of the most used query-based methods.
%The MMR family is represented by the original MMR, Expected n-call@k~\cite{Lim:2012}, and the Portfolio Theory~\cite{Wang:2009}.
Expected n-call@k adapts and extends MMR as a probabilistic model (Probabilistic Latent MMR).
The Portfolio Theory also extends MMR based on the idea of ranking
under uncertainty. As baseline, we used the straightforward idea of
combining all input documents into a single one, and then submit the
document to the single-document summarization method. Considering that
most coverage-based systems explore event information, 
we opted for not including them in this comparative
analysis.

% In practice, Expected n-call@k is equivalent to the MMR replacing the two cosine
% similarity distances by other distances. In particular, the cosine similarity
% between document and query is replaced by with the BM25 probabilistic relevance
% score~\cite{Robertson:1994:SEA} and the cosine distance between two documents is
% replaced by the TF-IDF similarity metric~\cite{salton1983introduction}.

To assess the informativeness of the summaries generated by our methods,
we used ROUGE-1 and ROUGE-2~\cite{lin:2004} on DUC 2007 and
TAC 2009 datasets. The main summarization task in DUC
2007\footnote{http://www-nlpir.nist.gov/projects/duc/duc2007/tasks.html}
is the generation of 250-word summaries of 45 clusters of 25 newswire
documents (from the AQUAINT corpus) and 4 human reference summaries. The
TAC 2009 Summarization
task\footnote{http://www.nist.gov/tac/2009/Summarization/} has 44 topic
clusters. Each topic has 2 sets of 10 news documents obtained from the
AQUAINT 2 corpus.%~\cite{Vorhees:2008aquaint2}.
There are 4 human
100-word reference summaries for each set, where the reference summaries
for the first set are query-oriented, and for the second set are update
summaries. In this work, we used the first set of reference summaries.
We evaluate the different models by generating summaries with 250 words.
We only present the best results.
%obtained with the best models for each dataset.

The used features include the bag-of-words model representation of the sentences
(TF-IDF), the key phrases and the query (obtained from the topics descriptions).
Including the query is a new extension to the \textsc{KP-Centrality} method,
which, in general, improved the results. We experimented with different numbers
of key phrases, obtaining the best results with 40 key phrases. To compare and
rank the sentences, we use several distance metrics, namely: Frac133 (generic
Minkowski distance, with $N=1.(3)$), Euclidean, Chebyshev, Manhattan, Minkowski,
the Jensen-Shannon Divergence, and the cosine similarity.
\begin{table*}
	\centering %\small
	\begin{tabular}{c c  c c  c c}
		\hline
				 &							  & \multicolumn{2}{c}{DUC 2007}   &  \multicolumn{2}{c}{TAC 2009}  \\
		Distance & Model                      & R1 & R2         & R1 & R2  \\
		\hline
		\hline
		frac133 & \multirow{ 2}{*}{baseline} & 0.3565 & 0.0744  & 0.4706 & 0.1268 \\
		cosine  &                            & 0.3406 & 0.0670  & 0.4746 & 0.1391 \\
		\hline
		frac133 & waterfall               & 0.3569          &  0.0765 & 0.4943 & 0.1441  \\
		frac133 & single-layer               & \textbf{0.3775} &  0.0882 & 0.4983 & 0.1526  \\
		cosine  & waterfall                  & 0.3701          & \textbf{0.0904}  & \textbf{0.5137} & \textbf{0.1693} \\
		cosine  & single-layer            & 0.3707          & 0.0822  & 0.4993  & 0.1590 \\

		frac133 & single-layer (shuffle)        & 0.3689  & 0.0807  & 0.5060 & 0.1483  \\
		cosine & waterfall (shuffle)             & 0.3626 & 0.0844  & 0.5107 & 0.1630 \\
		%frac133 & single-l. (TI) & 0.3650   & 0.5081\\
		%frac133 & waterfall (TI) & 0.3763   & 0.5129\\
		%cosine  & waterfall (TI) & 0.3766   & 0.4920\\
		\hline
		\multicolumn{2}{c}{MEAD}       & 0.3282 & 0.0765 & 0.4153 & 0.0845 \\
		\multicolumn{2}{c}{MMR}        & 0.3269 & 0.0780 & 0.3917 & 0.0801 \\
		\multicolumn{2}{c}{E.n-call@k} & 0.3209 & 0.0701 & 0.3873 & 0.0699 \\
		\multicolumn{2}{c}{Portfolio}  & 0.3595 & 0.0792 & 0.4292 & 0.0758 \\
		\multicolumn{2}{c}{LexRank}    & 0.2881 & 0.0534 & 0.3845 & 0.0623 \\
		\hline
	\end{tabular}\caption{ROUGE-1 (R1) and ROUGE-2 (R2) scores.\label{table:ResultsAll}}
\end{table*}
Table~\ref{table:ResultsAll} shows that the best results were obtained
by the proposed hierarchical models, in both datasets. Overal, the best
performing distance metric for our centrality-based method was the
cosine similarity and the best strategy for combining the information
was the waterfall approach, namely, in terms of ROUGE-2. In DUC 2007,
frac133 using the single-layer method achieved the best ROUGE-1 score,
although the difference for cosine is hardly noticeable. Single-layer
with frac133 shows a performance improvement of 0.0180 ROUGE-1 points
(relative performance improvement of 5.0\%) over the best of the other
systems, Portfolio, in DUC 2007, and of 0.0845 ROUGE-1 points (19.7\%
relative performance improvement) in TAC 2009. In terms of ROUGE-2, the
waterfall method using cosine achieved an improvement of 0.0112
(relative performance improvement of 14.1\%) over Portfolio, in DUC
2007, and of 0.0848 (relative performance improvement of 100.4\%) over
MEAD, the best performing of the reference systems using this metric, in
TAC 2009. Note that our baseline obtained results similar to the best
reference system in DUC 2007 and better results than all reference
systems in TAC 2009 (0.0454 ROUGE-1 points corresponding to a 10.6\%
relative performance improvement; 0.0546 ROUGE-2 points corresponding to
a 64.6\% relative performance improvement). The better results obtained
on the TAC 2009 dataset are due to the small size of the reference
summaries and to the fact that the documents sets to be summarized
contain topics with higher diversity of subtopics.

%can be justified by two
%reasons: the reference summaries are smaller (thus, selection errors
%%have an higher impact on the results) and document sets to be summarized
%contain topics with higher diversity of subtopics.
%
% Considering that the proposed mehtods when using temporal information achieve
% better results than all other methods, including the ours when not using
% temporal,
%

The shuffle results included in Table~\ref{table:ResultsAll} are
averages of 10 trials. They are lower than the other obtained using the
documents organized in chronological order. This suggests that the order
of the input documents is important to the summarization methods.

Figure~\ref{tb:ExampleOfSummary} shows an example of summary produced by our multi-document method. The figure also includes the respective reference summary for comparison.
\begin{figure}
	\centering \footnotesize
	\def\arraystretch{2}%
	\begin{tabular}{p{.9\columnwidth}} \hline
Generated Summary: \\ \hline	
President Bill Clinton
said Friday he will appeal a federal judge's ruling that struck
down a law giving the president the power to veto specific items
in bills passed by Congress.
The law, passed by Congress last year, allowed the president
for the first time to veto particular items in spending bills and
certain limited tax provisions passed by Congress.
Clinton said the funding that Congress has added to the bill is
excessive and threatened to veto some items by using the line-item
veto power.
The White House said that the president used his authority to
cancel projects that were not requested in the budget and would not
substantially improve the quality of life of military service
members.
Judge Thomas Hogan ruled that the law -- which gives the
president the power to strike items from tax and spending measures
without vetoing the entire bill -- violates the traditional
balance of powers between the various branches of government
"The Line-Item Veto Act is unconstitutional because it
impermissibly disrupts the balance of powers among the three
branches of government," said Thomas Hogan."
In its appeal, the Justice Department argues that the new
challengers also do not have standing to challenge the law, and
that in any case the law is in line with the historic relationship
between Congress and the president. \\ \hline
Reference summary:\\ \hline
Congress passed a law authorizing the line item veto (LIV) in 1996 accepting arguments that the measure would help preserve the integrity of federal spending by allowing the president to strike unnecessary spending and tax items from legislation thus encouraging the government to live within its means.  It was considered in line with the historic relationship between Congress and the president and would provide a tool for eliminating wasteful pork barrel spending while enlivening debate over the best use of funds.  It was argued that the LIV would represent presidential exercise of spending authority delegated by Congress.
President Clinton exercised the LIV on 82 items in 1997 saving \$1.9 billion in spending projected over five years.  The affected items were projects for specific localities, many in the area of military construction, which had been added to the president's budget by Congress.
The first court ruling on the LIV act was in U.S. District Court when in February 1998 it was ruled unconstitutional on the grounds that it violated the separation of powers.  The Department of Justice appealed that decision and in June 1998 the Supreme Court ruled the LIV act unconstitutional but on the grounds that it violated Article I, 7, Clause 2 (The "presentment clause") of the Constitution that establishes the process by which a bill becomes law.
President Clinton expressed his deep disappointment.\\
\hline
		\hline
	\end{tabular}  \caption{Example of summary produced by our summarizer and the reference summary Topic D0730G of DUC 2007}
\label{tb:ExampleOfSummary}
\end{figure}
\normalsize

\section{Conclusions and Future Work}
In this work, we explore two different approaches to extend a
single-document summarization method to multi-document summarization:
single-layer hierarchical and waterfall.

Experimental results show that the proposed approaches perform better
than previous state-of-the-art methods on standard datasets used to
evaluate this task. In general, the best performing approach is the
waterfall approach using the cosine similarity.  In fact, this
configuration achieves the best results on the TAC 2009 dataset,
considering both ROUGE-1 and ROUGE-2 metrics, and, although not
achieving the best results in the DUC 2007 dataset, in terms of ROUGE-1,
it also achieves a performance improvement over Portfolio of 0.0106
ROUGE-1 points (relative performance improvement of 3\%).

In future work, we aim to adapt the proposed multi-document
summarization method to perform abstractive summarization. 

\section*{Acknowledgments}

This work has been partially supported by national funds through
Funda\c{c}\~{a}o para a Ci\^{e}ncia e a Tecnologia (FCT) with reference
UID/CEC/50021/2013, the grant numbers SFRH/BD/33769/2009 and
CMUP-EPB/TIC/0026/2013. The authors would also like to thank Eduard
Hovy, Isabel Trancoso, Ricardo Baeza-Yates, and the anonymous reviewers
for fruitful comments.

\bibliographystyle{acl}
\bibliography{sigirrsp098} 

\end{document}